\documentclass[onecolumn]{article}%
\usepackage{amssymb}
\usepackage{amsfonts}
\usepackage{amsmath}
\usepackage{graphicx}
\usepackage{xcolor}%
\providecommand{\U}[1]{\protect\rule{.1in}{.1in}}
 \arraycolsep=1.5pt 

\ifx\pdfoutput\relax\let\pdfoutput=\undefined\fi
\newcount\msipdfoutput
\ifx\pdfoutput\undefined\else
\ifcase\pdfoutput\else
\ifx\paperwidth\undefined\else
\ifdim\paperheight=0pt\relax\else\pdfpageheight\paperheight\fi
\ifdim\paperwidth=0pt\relax\else\pdfpagewidth\paperwidth\fi
\fi\fi\fi
\begin{document}

\title{\textbf{{\Large {Similarity reductions and new nonlinear exact solutions for
the 2D incompressible Euler equations} }}}
\author{\textsc{Engui Fan\thanks{ E-mail address: faneg@fudan.edu.cn}}\\\textit{School of Mathematical Sciences},\\[-0.12in] \textit{Shanghai Center for Mathematical Sciences}\\[-0.12in] \textit{and Key Laboratory of Mathematics for Nonlinear Science},\\[-0.12in]\textit{Fudan University, Shanghai 200433, P.R. China}
\and M\textsc{anwai Yuen\thanks{Corresponding author and e-mail address:
nevetsyuen@hotmail.com}}\\\textit{Department of Mathematics and Information Technology,}\\[-0.12in]\textit{The Hong Kong Institute of Education,}\\[-0.12in]\textit{10 Po Ling Road, Tai Po, New Territories, Hong Kong}}
\date{Revised 29-Nov-2013}
\maketitle

\begin{abstract}
For the 2D and 3D Euler equations, their existing exact solutions are often in
linear form with respect to variables $x,y,z$. In this paper, the
Clarkson-Kruskal reduction method is applied to reduce the 2D incompressible
Euler equations to a system of completely solvable ordinary equations, from
which several novel nonlinear exact solutions with respect to the variables
$x$ and $y$ are found. \newline MSC: 35Q31, 35C05, 76B03, 76M60\newline Key
Words: Incompressible Euler equations, the Clarkson-Kruskal method, similarity
reductions, nonlinear exact solutions.

\end{abstract}

\section{Introduction}

The motion of inviscid, incompressible-ideal fluid is governed by the
following equations
\begin{equation}
\label{Euler1}%
\begin{split}
&  \mathrm{div} u=0,\\
&  u_{t}+(u\cdot\triangledown)u+\triangledown p=0,
\end{split}
\end{equation}
which was first obtained by Euler. Here $u=(u_{1}, u_{2}, u_{3})$ are the
components of the three dimensional velocity field and $p$ the pressure of the
fluid at a position $(x,y,z)$. The Euler equations (\ref{Euler1}) are not in
Hamiltonian form owing to the lack of an equation explicitly governing the
time evolution of the pressure. Arnold's strategy to obviate this difficulty
was to project each term of the system (\ref{Euler1}) onto a canonically
chosen divergence-free representative, thereby eliminating the pressure terms
at the expense of introducing a nonlocal operator on the nonlinear terms
\cite{Ar}. Here the more conventional procedure of simply taking the curl of
(\ref{Euler1}) performs the same function, leading to a vorticity equation
\begin{equation}
\label{Euler2}%
\begin{split}
&  \partial_{t}\Omega+(u\cdot\triangledown)\Omega-(\Omega\cdot\triangledown
)u=0.
\end{split}
\end{equation}
Kato showed that the mild solutions of the 2D Navier-Stokes equations approach
the 2D Euler equations \cite{Kato}. Marsden and Ratiu carried out extensive
studies on the sympletic structure of 2D Euler model \cite{MR}. Friedlander
and Vishik constructed a Lax pair for Euler equations in the Lagrangian
coordinates \cite{FV}. Recently, Li found weak Lax pairs for 2D and 3D Euler
equations (\ref{Euler2}) in the vorticity form \cite{Li1, Li2}. Lou and Li
proposed Backlund transformation, Darboux transformation and exact solutions
for the 2D Euler equations in the vorticity form \cite{Lou1, Lou2}. Interest
in self-similar solutions can be traced back to the work of Sedov and
Barenblatt \cite{34,5,6}. In 1965, Arnold first introduced the famous
Arnold-Beltrami-Childress (ABC) flow \cite{4}. The solutions exhibit
interesting local behavior with infinite energy. Zelik's work gives details of
the existence of weak solutions for the unbounded domain \cite{41,42}. The
exact solutions to the infinite energy of the systems can be regionally
applied to understand the great complexity that exists in turbulence
\cite{28}. Makino obtained the first radial solutions to the Euler and
Navier-Stokes equations in 1993, using the separation method \cite{31}. In
2011, Yuen obtained a class of exact and rotational solutions for the 3D Euler
equations \cite{Yuen1}. In 2012, Yuen constructed exact solutions with
elliptical symmetry by using the new characteristic method \cite{Yuen2, Yuen3}.

We notice that these existing exact solutions $u_{1}, u_{2}, u_{3}$ for the 2D
and 3D Euler equations are often in linear form with respect to special
variables $x, y, z$. So it is a natural idea to search for their novel
nonlinear exact solutions with respect to variables $x, y$ and $y$. One of
best methods to achieve this goal may be the reduction method developed by the
Clarkson and Kruskal \cite{Clarkson}, because their basic idea is to seek a
reduction of a given partial differential equation in the form
\begin{equation}
\label{reduction1}%
\begin{split}
&  u=\alpha(x,t)+\beta(x,t)W(z(x,t)),
\end{split}
\end{equation}
which should cover most general form of the known solutions for the Euler
equations. Substituting (\ref{reduction1}) into the partial differential
equation and demanding that the result be ordinary differential equation for
$W$ of $z$ through certain constraints on its derivatives the ratios of their
coefficients of different derivatives and powers being functions of $z$ only.
The ordinary differential equation may be a special or solvable equation. The
unusual characteristic of this method is that it does not use Lie group theory
originally developed by Lie \cite{Lie}. Though the Lie group method is
entirely algorithmic, it often involves a large amount of tedious algebra and
auxiliary calculations which are virtually unmanageable manually.

Based on the above consideration, as illustrative examples in this paper, we
would like to apply the Clarkson-Kruskal reduction method to find novel
nonlinear solutions for the 2D case of non-vorticity Euler equations
(\ref{Euler1}), which can be written in the scalar form
\begin{equation}%
\begin{split}
&  \frac{\partial u_{1}}{\partial x}+\frac{\partial u_{2}}{\partial y}=0,\\
&  \frac{\partial u_{1}}{\partial t}+u_{1}\frac{\partial u_{1}}{\partial
x}+u_{2}\frac{\partial u_{1}}{\partial y}+\frac{\partial p}{\partial x}=0,\\
&  \frac{\partial u_{2}}{\partial t}+u_{1}\frac{\partial u_{2}}{\partial
x}+u_{2}\frac{\partial u_{2}}{\partial y}+\frac{\partial p}{\partial y}=0,
\end{split}
\label{Euler3}%
\end{equation}
in which $u_{1}=u_{1}(x,y,t),u_{2}=u_{2}(x,y,t)$ are the velocity of fluid,
and $p=p(x,y,t)$ is the pressure. It will be shown that our solutions are
different from the existing linear form of solutions with respect to variables
$x$ and $y$ for the 2D incompressible Euler equations (\ref{Euler3}).

This paper is organized as follows. In section 2, we construct the similarity
reductions of Euler equations (\ref{Euler3}) by using the Clarkson-Kruskal
reduction method. In section 3, as special applications of the similarity
reductions, we find some interesting nonlinear solutions of the Euler
equations (\ref{Euler3}).

\section{The similarity reductions}

In this section, we seek similarity reduction of the 2D Euler equations
(\ref{Euler3}) in the form%

\begin{equation}
\label{2.1}%
\begin{split}
&  u_{1}=\alpha(x,y,t)+\beta(x,y,t)W(z(x,y,t)),\\
&  u_{2}=\xi(x,y,t)+\eta(x,y,t)Q(z(x,y,t)),\\
&  p=p(x,y,t),
\end{split}
\end{equation}
where $\alpha(x,y,t), \beta(x,y,t),\xi(x,y,t), \eta(x,y,t), p(x,y,t)$ and
$z(x,y,t)$ are functions to be determined. Since $p(x,y,t)$ is a linear
function in (\ref{Euler3}), its most general form for a similarity is itself.
Here we take it to be a determined function such that final reduced ordinary
equations are compatible.

Substituting (\ref{2.1}) into (\ref{Euler3}), and collecting coefficients of
monomials of $W, Q$ and their derivatives yields
\begin{equation}
\label{eq1}%
\begin{split}
&  \beta z_{x}W^{\prime}+\eta z_{y} Q^{\prime}+\beta_{x}W+\eta_{y}Q+\alpha
_{x}+\xi_{y}=0,
\end{split}
\end{equation}
\begin{equation}
\label{eq2}%
\begin{split}
&  \beta^{2}z_{x}WW^{\prime}+(\beta z_{t}+\alpha\beta z_{x}+\xi\beta
z_{y})W^{\prime}+\eta\beta z_{y}QW^{\prime}+\beta\beta_{x}W^{2},\\
&  +(\beta_{t}+\alpha\beta_{x}+\alpha_{x}\beta+\xi\beta_{y})W+\eta\beta
_{y}QW+\eta\alpha_{y}Q\\
&  + \alpha_{t}+\alpha\alpha_{x}+\xi\alpha_{y}+p_{x}=0
\end{split}
\end{equation}
and
\begin{equation}
\label{eq3}%
\begin{split}
&  \eta^{2}z_{y}QQ^{\prime}+(\eta z_{t}+\alpha\eta z_{x}+\xi\eta
z_{y})Q^{\prime}+\eta\beta z_{x}WQ^{\prime}+\eta\eta_{y}Q^{2},\\
&  +(\eta_{t}+\alpha\eta_{x}+\xi\eta_{y}+\eta\xi_{y})Q+\beta\eta_{x}%
QW+\beta\xi_{x}W\\
&  + \xi_{t}+\alpha\xi_{x}+\xi\xi_{y}+p_{y}=0
\end{split}
\end{equation}
where $^{\prime}:=\frac{d}{dz}$. In order to make these equations be a system
of ordinary differential equations for $W$ and $Q$, the ratios of their
coefficients of different derivatives and powers have to be functions of $z$
only. That is, the following equations should be satisfied
\begin{align}
&  \eta z_{y}=\beta z_{x}\Gamma_{1}(z),\label{eq8}\\
&  \beta_{x}=\beta z_{x}\Gamma_{2}(z),\label{eq9}\\
&  \eta_{y}=\beta z_{x}\Gamma_{3}(z),\label{eq10}\\
&  \alpha_{x}+\xi_{y}=\beta z_{x}\Gamma_{4}(z),\label{eq11}\\
&  \beta z_{t}+\alpha\beta z_{x}+\xi\beta z_{y}=\beta^{2} z_{x}\Gamma
_{5}(z),\label{eq12}\\
&  \eta\beta z_{y}=\beta^{2} z_{x}\Gamma_{6}(z),\label{eq13}\\
&  \beta\beta_{x}=\beta^{2} z_{x}\Gamma_{7}(z),\label{eq14}\\
&  \beta_{t}+\alpha\beta_{x}+\alpha_{x}\beta+\xi\beta_{y}=\beta^{2}
z_{x}\Gamma_{8}(z),\label{eq15}\\
&  \eta\beta_{y}=\beta^{2} z_{x}\Gamma_{9}(z),\label{eq16}\\
&  \eta\alpha_{y}=\beta^{2} z_{x}\Gamma_{10}(z),\label{eq17}\\
&  \alpha_{t}+\alpha\alpha_{x}+\xi\alpha_{y}+p_{x}=\beta^{2} z_{x}\Gamma
_{11}(z)\label{eq18}\\
&  \eta z_{t}+\alpha\eta z_{x}+\xi\eta z_{y}=\eta^{2} z_{y}\Gamma
_{12}(z),\label{eq19}\\
&  \eta\beta z_{x}=\eta^{2} z_{y}\Gamma_{13}(z),\label{eq20}\\
&  \eta\eta_{y}=\eta^{2} z_{y}\Gamma_{14}(z),\label{eq21}\\
&  \eta_{t}+\alpha\eta_{x}+\xi\eta_{y}+\eta\xi_{y}=\eta^{2} z_{y}\Gamma
_{15}(z),\label{eq22}\\
&  \beta\eta_{x}=\eta^{2} z_{y}\Gamma_{16}(z),\label{eq23}\\
&  \beta\xi_{x}=\eta^{2} z_{y}\Gamma_{17}(z),\label{eq24}\\
&  \xi_{t}+\alpha\xi_{x}+\xi\xi_{y}+p_{y}=\eta^{2} z_{y}\Gamma_{18}(z),
\label{eq25}%
\end{align}
where $\Gamma_{i}(z)\ (i=1,\cdots,18)$ are some arbitrary functions of $z$
that will be determined later. In the determination of functions $\alpha,
\beta, \xi, \eta$ and $z$, there exist some freedoms without loss of generality.

\textit{Rule 1.} If $\alpha$ has the form $\alpha=\alpha_{0}+\beta\Omega(z)$,
then we can take $\Omega(z)=0$ ( by taking $W\rightarrow W-\Omega)$).

\textit{Rule 2.} If $\beta$ has the form $\beta=\beta_{0} \Omega(z)$, then we
can take $\Omega(z)=1$ ( by taking $\Omega\rightarrow W/\Omega$).

\textit{Rule 3.} If $z$ is determined by an equation of the form
$\Omega(z)=z_{0}$, then we can take $\Omega(z)=z$ ( by taking $z\rightarrow
\Omega^{-1}(z)$). The other functions $\xi$ and $\eta$ can also be determined
in a similar way as above.

We shall now proceed to determine the general similarity reductions of the 2D
Euler equations (\ref{Euler3}) using this method. From the constraint equation
(\ref{eq9}), we have
\begin{equation}
\beta_{x}/\beta=z_{x}\Gamma_{2}(z), \label{2.27}%
\end{equation}
which upon integration gives
\begin{equation}
\beta=\beta_{0}(y,t)\exp(\Gamma_{2}(z)), \label{2.27}%
\end{equation}
where $\beta_{0}(y,t)$ is a function of integration. Hence, using freedom
mentioned in rule 2 above, we choose $\Gamma_{2}(z)=0$ in (\ref{2.27}) to
obtain
\begin{equation}
\beta=\beta_{0}(y,t). \label{2.28}%
\end{equation}
For the simplicity of reduction, we just consider $\beta$ being constant and
take
\begin{equation}
\beta=1. \label{eq29}%
\end{equation}

In a similar way to deal with $\beta$, from constraint equation (\ref{eq21})
and rule 2, we may take
\begin{equation}
\Gamma_{14}(z)=0, \ \ \eta=1. \label{1.29}%
\end{equation}
Then it follows from (\ref{eq8}), (\ref{eq29}) and (\ref{1.29}) that
\[
\Gamma_{1}(z)=1.
\]

By using (\ref{eq29}) and (\ref{1.29}), it follow (\ref{eq10}), (\ref{eq14}),
(\ref{eq21}), and (\ref{eq23}) that
\begin{equation}
\label{2.29}\Gamma_{3}(z)=\Gamma_{7}(z)=\Gamma_{14}(z)=\Gamma_{16}(z)=0.
\end{equation}
Substituting (\ref{eq29}) into (\ref{eq15}) and integrating, we have
\begin{equation}
\alpha=\theta_{1}(y,t)+\Gamma_{8}(z), \label{2.29}%
\end{equation}
which, by the rule 1, leads to
\begin{equation}
\Gamma_{8}(z)=0,\ \ \ \alpha=\theta_{1}(y,t). \label{eq33}%
\end{equation}
In the similar way, we can obtain from (\ref{eq22}) that
\begin{equation}
\Gamma_{15}(z)=0,\ \ \ \xi=\theta_{2}(x,t). \label{eq34}%
\end{equation}

Substituting (\ref{eq33}) and (\ref{eq34}) into (\ref{eq11}) gives
\begin{equation}
\Gamma_{4}(z)=0. \label{2.32}%
\end{equation}
It is easily seen from (\ref{eq17}) and (\ref{eq33}) that
\begin{equation}
\theta_{1,y}(y,t)x+\theta_{3}(y,t)=\Gamma_{10}(z), \label{2.32}%
\end{equation}
which implies that
\begin{equation}
\Gamma_{10}(z)=1,\ \ z=\theta_{1,y}x+\theta_{3}(y,t) \label{eq37}%
\end{equation}
combination with rule 3. Substituting (\ref{eq37}) into (\ref{eq20}) yields
\begin{equation}
\Gamma_{13}(z)=1,\ \ \theta_{1,yy}=0,\ \ \theta_{1,y}=\theta_{3,y},
\label{2.32}%
\end{equation}
which leads to
\begin{equation}
\theta_{1}=\theta_{3}=f(t)y+g(t),\ \ \alpha(y,t)=f(t)y+g(t) \label{eq39}%
\end{equation}
where $f(t)$ and $g(t)$ are arbitrary functions of $t$.

From (\ref{eq37}) and (\ref{eq39}), we can write the variable $z$ in the form
\begin{equation}
\label{eq40}z=f(t)x+f(t)y+g(t).
\end{equation}
Substituting it into (\ref{eq13}) gives
\begin{equation}
\label{2.32}\Gamma_{6}(z)=1.
\end{equation}

Substituting (\ref{eq34}) and (\ref{eq37}) into (\ref{eq24}) gives
\begin{equation}
\label{eq42}\Gamma_{17}(z)=1, \ \ \ \xi=\theta_{2}=f(t)x+h(t).
\end{equation}

On use of (\ref{eq29}) and (\ref{eq42}), the equations (\ref{eq18}) and
(\ref{eq25}) can be written as
\begin{align}
&  f^{2}(t)x+f^{\prime}(t)y+(g^{\prime}(t)+f(t)h(t))+p_{x}=z_{x}\Gamma
_{11}(z),\label{eq43}\\
&  f^{2}(t)y+f^{\prime}(t)x+(h^{\prime}(t)+f(t)g(t))+p_{y}=z_{y}\Gamma
_{18}(z). \label{eq44}%
\end{align}
Integrating equations (\ref{eq43}) and (\ref{eq44}) with respect to variable
$x$ and $y$, again using rule 1, we obtain that
\begin{align}
\label{2.32} &  \Gamma_{11}(z)=\Gamma_{18}(z)=0, \ \ h(t)=g(t),\\
&  p=-\frac{1}{2}f^{2}(t)(x^{2}+y^{2})-f^{\prime}(t)xy-(g^{\prime
}(t)+f(t)g(t))(x+y).
\end{align}

Substituting (\ref{eq39}), (\ref{eq40}) and (\ref{eq42}) into the equations
(\ref{eq12}) and (\ref{eq19}) gives
\begin{equation}
f^{\prime}(t)(x+y)+g^{\prime}(t)+2f(t)g(t)=f(t)\Gamma_{5}(z)=f(t)\Gamma
_{12}(z). \label{eq47}%
\end{equation}
Observing $z=f(t)x+f(t)y+g(t)$ in (\ref{eq40}) and the left-hand side of
equation (\ref{eq47}) is linear in $x$ and $y$, consequently, the functions
$\Gamma_{5}(z)$ and $\Gamma_{12}(z)$ should be linear function of $z$, that
is,
\begin{equation}
\Gamma_{5}(z)=\Gamma_{12}(z)=c_{1}z+c_{2} \label{2.32}%
\end{equation}
with $c_{1}$, $c_{2}$ being arbitrary constants. Balancing coefficients of
powers of $x$ and $y$ in (\ref{eq47}) implies that $f(t)$ and $g(t)$ satisfy
\begin{equation}
\label{eq49}%
\begin{split}
&  f^{\prime}(t)=(c_{1}-1)f^{2}(t),\\
&  g^{\prime}(t)=(c_{1}-2)f(t)g(t)+c_{2}f(t).
\end{split}
\end{equation}

According to the values of obtained functions $\Gamma_{i}(z)\ (i=1,\cdots
,18)$, the 2D Euler equations (\ref{Euler3}) are reduced to a system of the
ordinary differential equations with respect to $W(z)$ and $Q(z)$
\begin{align}
&  W^{\prime}+Q^{\prime}=0,\nonumber\\
&  WW^{\prime}+(c_{1}z+c_{2})W^{\prime}+QW^{\prime}+Q=0,\label{eq51}\\
&  QQ^{\prime}+(c_{1}z+c_{2})Q^{\prime}+WQ^{\prime}+W=0.\nonumber
\end{align}

Finally, we conclude that the general similarity reduction for 2D Euler
equations (\ref{Euler3}) is given by
\begin{equation}
\label{eq51}%
\begin{split}
&  u_{1}=f(t)y+g(t)+W(z),\\
&  u_{2}=f(t)x+g(t)+Q(z),\\
&  p=-\frac{1}{2}f(t)^{2}(x^{2}+y^{2})-f^{\prime}(t)xy-[g^{\prime
}(t)+f(t)g(t)](x+y),\\
&  z=f(t)(x+y)+g(t),
\end{split}
\end{equation}
where $f(t)$ and $g(t)$ satisfy equation (\ref{eq49}), and $W(z)$, $Q(z)$
satisfy equation (51).

\section{Nonlinear exact solutions}

In this section, we discuss nonlinear exact solutions of the 2D Euler
equations (\ref{Euler3}) from the similarity reduction (\ref{eq51}). It is
important to deal with completely solvable system of ordinary equations (50)
and (51).

For different cases of parameters $c_{1}$ and $c_{2}$, the solutions of
ordinary differential equations (\ref{eq49}) are known as
\begin{equation}
f(t)=c_{3}=\mathrm{constant},\ \ g(t)=e^{-c_{3}t}+c_{2},\ \ \mathrm{for}%
\ \ c_{1}=1, \label{eq52}%
\end{equation}%
\begin{equation}
f(t)=-t^{-1},\ \ \ g(t)=-c_{2}\ln t,\ \ \mathrm{for}\ \ c_{1}=2 \label{eq53}%
\end{equation}
and
\begin{equation}
f(t)=\frac{1}{(1-c_{1})t},\ \ \ g(t)=t^{-\gamma}-\frac{c_{2}}{c_{1}%
-2},\ \ \gamma=\frac{c_{1}-2}{c_{1}-1},\ \ \mathrm{for}\ \ c_{1}\not =1,2.
\label{eq59}%
\end{equation}

And equations (51) has three classes of solutions for different parameters
$c_{1}$ and $c_{2}$
\begin{equation}
W=-Q=(z+c_{2}/c_{1})^{1/c_{1}},\ \ \mathrm{for}\ \ c_{1}\not =0. \label{eq54}%
\end{equation}%
\begin{equation}
W=-Q=e^{z/c_{2}},\ \ \mathrm{for}\ \ c_{1}=0,\ c_{2}\not =0. \label{eq55}%
\end{equation}
and
\begin{equation}
W=Q=0,\ \ \mathrm{for}\ \ c_{1}=c_{2}=0. \label{eq56}%
\end{equation}

\textbf{Case 1:} $c_{1}=c_{2}=0$. In this case, it follows from (\ref{eq51}),
(\ref{eq52}) and (\ref{eq56}) that the solution of the 2D Euler equations
(\ref{Euler3}) is given by
\begin{align}
&  u_{1}=t^{-1}y+t^{-2},\nonumber\\
&  u_{2}=t^{-1}x+t^{-2},\\
&  p=-t^{-2}(x-y)^{2}/2+t^{-3}(x+y),\nonumber
\end{align}
which is a rational solution and $u_{1}$, $u_{2}$ are linear with respect to a
single spacial variable.

\textbf{Case 2: } $c_{1}=0,\ \ c_{2}\not =0$. In this case, it follows from
(\ref{eq51}), (\ref{eq52}) and (\ref{eq55}) that the solution of the 2D Euler
equations (\ref{Euler3}) is given by
\begin{align}
&  u_{1}=t^{-1}y+t^{-2}+c_{2}/2+\exp\left[  c_{2}^{-1}t^{-2}%
(tx+ty+1)+1/2\right]  ,\nonumber\\
&  u_{2}=t^{-1}x+t^{-2}+c_{2}/2-\exp\left[  c_{2}^{-1}t^{-2}%
(tx+ty+1)+1/2\right]  ,\label{eq59}\\
&  p=-t^{-2}(x-y)^{2}/2+(t^{-3}-c_{2}t^{-1}/2)(x+y),\nonumber
\end{align}
in which $u_{1}$ and $u_{2}$ for variables $x$ and $y$ are linear. These two
kinds of solutions were never obtained by Yuen \cite{Yuen1}.

\textbf{Case 3:} $c_{1}=1$. In this case, it follows from (\ref{eq51}),
(\ref{eq52}) and (\ref{eq54}) that the solution of the 2D Euler equations
(\ref{Euler3}) is given by
\begin{align}
&  u_{1}=c_{3}(x+2y)+2e^{-c_{3}t}+3c_{2},\nonumber\\
&  u_{2}=-c_{3}y-c_{2},\\
&  p=-\frac{1}{2}c_{3}^{2}(x^{2}+y^{2})-c_{2}c_{3}(x+y).\nonumber
\end{align}

\textbf{Case 4:} $c_{1}=2$. In this case, it follows from (\ref{eq51}),
(\ref{eq53}) and (\ref{eq54}) that the solution of the 2D Euler equations
(\ref{Euler3}) is given by
\begin{align}
&  u_{1}=-t^{-1}y-c_{2}\ln t+[-t^{-1}(x+y)-c_{2}\ln t+c_{2}/2]^{1/2}%
,\nonumber\\[4pt]
&  u_{2}=-t^{-1}x-c_{2}\ln t-[-t^{-1}(x+y)-c_{2}\ln t+c_{2}/2]^{1/2}%
,\label{eq61}\\[4pt]
&  p=-\frac{1}{2}t^{-2}(x+y)^{2}+c_{2}t^{-1}(1-\ln t)(x+y).\nonumber
\end{align}

\textbf{Case 5: } $c_{1}\not =0,1,2$. In this case, it follows from
(\ref{eq51}), (\ref{eq59}) and (\ref{eq54}) that the solution of the 2D Euler
equations (\ref{Euler3}) is given by
\begin{align}
&  u_{1}={yt^{-1}}{(1-c_{1})^{-1}}+t^{-\gamma}-{c_{2}}{(c_{1}-2)^{-1}}+\left[
-{(x+y)t^{-1}}{(1-c_{1})^{-1}}+t^{-\gamma}-{2c_{2}}{c_{1}^{-1}(c_{1}-2)^{-1}%
}\right]  ^{1/c_{1}},\nonumber\\[4pt]
&  u_{2}={xt^{-1}}{(1-c_{1})^{-1}}+t^{-\gamma}-{c_{2}}{(c_{1}-2)^{-1}}-\left[
-{(x+y)t^{-1}}{(1-c_{1})^{-1}}+t^{-\gamma}+{2c_{2}}{c_{1}^{-1}(c_{1}-2)^{-1}%
}\right]  ^{1/c_{1}},\label{eq62}\\[4pt]
&  p=-{(x^{2}+y^{2})t^{-2}}{(c_{1}-1)^{-2}}/2-{xyt^{-2}}{(1-c_{1})^{-1}%
}+\left(  -\gamma t^{-\gamma}-{c_{2}t^{-1}}{(1-c_{1})^{-1}(c_{1}-2)^{-1}%
}\right)  (x+y).\nonumber
\end{align}
It is obvious that $u_{1}$ and $u_{2}$ in (\ref{eq59}), (\ref{eq61}) and
(\ref{eq62}) are not linear with respect to variables $x$ and $y$. These
solutions, to the best of our knowledge, should be previously unknown. Here we
should remark that, as usual, the 2D Euler equations should be reduced to a
system of (1+1)-dimensional partial different equations. Here we take full
advantage of the divergence equation $\mathrm{div}u=0$ in the incompressible
case to directly reduce the 2D Euler equations (\ref{Euler3}) to a system of
solvable ordinary equations. Consequently, we are able to construct several
novel exact solutions for the incompressible 2D Euler equations. This method
also can be applied to the incompressible 3D Euler equations. In addition, Lou
and Li found solitary wave solutions for the 2D Euler equations in the
vorticity form through their Lax and Darboux transformation \cite{Lou2}. But
solitary wave solutions for general 2D and 3D Euler equations still have not
been found, partly due to the fact that their Lax pairs are unknown at present.

\section*{Acknowledgments}

The work described in this paper was partially supported by the research grant
RG 53/2012-2013R from the Hong Kong Institute of Education, the National
Science Foundation of China (Project No. 11271079), Doctoral Programs
Foundation of the Ministry of Education of China.

\end{document}